\documentclass{aa}
\usepackage{graphicx}
\usepackage{txfonts}
\usepackage{natbib}
\bibpunct{(}{)}{;}{a}{}{,}

\begin{document}
\title{The remarkable light and colour variability of Small Magellanic Cloud Be stars}
\titlerunning{Magellanic cloud Be stars}
\author{        W.J. de Wit\inst{1} \and 
                H.J.G.L.M. Lamers\inst{2,3} \and 
                J.B. Marquette\inst{4} \and
                J.P. Beaulieu\inst{4}}  
 
\offprints{     dewit@obs.ujf-grenoble.fr}
\institute{     Laboratoire d'Astrophysique de Grenoble, Universit\'{e} Joseph
                Fourier, BP 53, 38041 Grenoble Cedex 9, France \and
                Astronomical Institute, Utrecht University, Princetonplein 5,
                3584 CC Utrecht, The Netherlands\and 
                SRON Laboratory for Space Research, Sorbonnelaan 2, 3584 CA Utrecht, The Netherlands\and
                Institut d'Astrophysique de Paris, 98$^{\rm bis}$ Boulevard Arago, 
                75014 Paris, France                                    
}
\date{Received date; accepted date}
\abstract
{}
{To test if the irregularly variable optical emission of hot stars in the Small Magellanic
  Cloud is due to variable amounts of Brehmstrahlung by ionized material in the
  circumstellar environment. The light curve variability of this sample of hot
  stars ($\sim 200$ objects) is described by Mennickent et al. (2002).}
{(1) Probing the relation between flux excess in the optical and in the 
  near-infrared $J$-band using EROS light curves (time span of 5 years) and
  2MASS photometry. (2) Analyzing the relation
  between colour and magnitude of the variability by means of an
  analytical, time-dependent model of emission from ionized material in an
  outflowing circumstellar disk.} 
{(1) A correlation between optical and near-IR flux excess is found. (2) A
  bi-valued relation between excess colour and magnitude for $\sim 100$ objects
  is discovered. It results in a loop structure in a Colour-Magnitude
  diagram. Significantly, this loop is traversed in a clockwise sense for
  $\sim90\%$ of the stars, and anti-clockwise for the remainder. (3) A simple analytical model of an
  outflowing disk and a variable mass loss rate is capable of
  reproducing observed light curves and the bi-valued colour-magnitude relation.}
 {The optical variability of a large fraction of the hot stars is due to
 variations in the amount of Brehmstrahlung. The circumstellar envelope can
 contribute up to 50\% of the total optical flux. This variability can be
 interpreted as due to an outflowing ionized CS disk, that evolves into a ring
 structure. The observed bi-valued colour-magnitude relation is an optical depth
 effect at the various wavelengths of interest. The loop is traversed clockwise
 by outflowing matter, but anti-clockwise by infalling material. As the material
 is generally outflowing, the CS environment of the star is progressively cleared once the
 star stops losing mass. This study makes use of public OGLE, MACHO and 2MASS
 data.}
\maketitle
\section{Introduction}
\label{parintro}
An important contribution to observational astronomy and in particular to the
studies of variable stars was made in the nineties by the microlensing
experiments: EROS (e.g. Aubourg et al. 1995)\nocite{1995A&A...301....1A}, MACHO
(e.g. Alcock et al. 1993)\nocite{1993Natur.365..621A}, OGLE (e.g. Udalski et
al. 1997)\nocite{1997AcA....47..319U}. The projects produced an enormous
quantity of photometric data for millions of stars in the Galactic Bulge and the
Magellanic Clouds covering nearly a decade of continuous observations. Among the
interesting discoveries reported are the Magellanic Clouds' blue variable
stars. Two papers in 2002 by Keller et al. (MACHO)\nocite{2002AJ....124.2039K}
and Mennickent et al. (OGLE)\nocite{2002A&A...393..887M} reported the huge
variety of photometric variability displayed by hundreds of blue stars in the
Large and Small Magellanic Cloud, respectively. Both papers type the various
light curves according to their morphology. A statistical comparison of various
characteristic quantities of the variability between LMC and SMC blue variable
stars resulted in pronounced differences for phenomenologically similar light
curves (Sabogal et al. 2005). Keller et al. (2002; hereafter K02) examine the
presence of Balmer line emission in a subset of their sample and find a detection
rate of 91\%. They conclude that the Large Magellanic Cloud population of blue
variable stars is consistent with the Be star phenomenon (see also Lamers et
al. 1999\nocite{1999A&A...341..827L} and de Wit et al. 2003\nocite{2003A&A...410..199D}).

In the present work, we focus on the phenomenal optical brightness and colour
variability of blue variable stars in the Small Magellanic Cloud. The stars
and their light curves were first described by Mennickent et al. (2002; hereafter
M02).  About 7\% of the SMC Be stars was actually found by M02 to have
well-defined {\it periodically} varying light curves (see Mennickent et
al. 2003\nocite{2003ipc..conf...89M}). These stars are not considered here. Instead we focus
on the {\it irregular} photometric variability displayed by the majority of the
Be stars on time scales of 1 to $~1000$ days and amplitudes up to
$1^{m}$. Given that the observed Balmer line emission variability correlates
with the photometric variability for similar LMC blue variables (K02) and the
fact that generally the stars have redder colours at brighter phases, bound-free
and free-free (bf-ff) emission originating in an ionized circumstellar envelope (CSE) is a
strong candidate for the observed optical variability. This is generally the case for
Galactic stars showing the Be phenomenon. A correlation between near-infrared
(near-IR) flux excess and optical flux excess should therefore exist for the SMC Be
stars.  Such a correlation was already shown to exist for Galactic Be stars by
Dachs \& Wamsteker (1982). We proceed therefore by cross-correlating the SMC
blue variable stars discovered and catalogued by M02 with the near-IR 2MASS catalogues.

The paper is organized as follows. In Sect.\,\ref{defi} we specify the way the
sample of blue variable SMC stars used in this contribution is defined. We note that in the
following we refer to the stars simply by Be stars. In Sect.\,\ref{feon} 
the presentation of the cross-correlation between 2MASS and the sample stars follows.
Then in Sect.\,\ref{colour} we describe the remarkable change in optical colour 
as a function of the brightness. This behaviour is modeled in Sect.\,\ref{model}. We discuss the results
and conclude in Sects.\,\ref{disc} and \ref{concl}.

\section{Data and completeness of the sample}
\label{defi}
\begin{figure}
\includegraphics[height=8cm,width=7cm,angle=90]{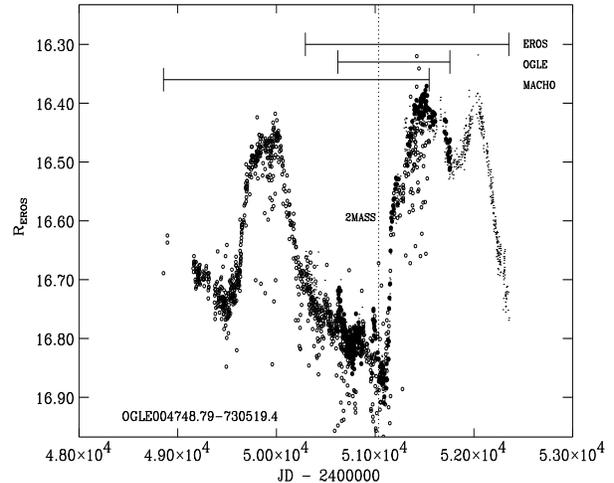}
 \caption{Light curves produced by three microlensing projects for the
   Small Magellanic Cloud Be star OGLE\,004748.79-730519.4. 
   Open circles are MACHO's red filter $R_{\rm M}$ data, filled circles are OGLE
   $I$-band data, small dots are EROS red filter $R_{\rm E}$ data. We have
   indicated the covered epoch of each project at the top. The 2MASS 
   epoch of observation is indicated by the 
   vertical dotted line and corresponds here to light minimum. In this
   particular case, $\Delta R$ (see Fig.\,\ref{ampl}) is nearly zero.}
\label{EOM}
\end{figure}

To investigate if the optical variability correlates with the near-IR flux, we need
to determine (1) the minimum light of each program star, supposedly corresponding to the
photospheric emission, and (2) the optical brightness of each object at
the epoch of the 2MASS near-IR observation. The type and time scale of the displayed
variability requires the longest period of observations for a proper
analysis. The light variability of the program stars maybe characterized as erratic
and an example is given in Fig.\,\ref{EOM}.

In this paper we use EROS\,II microlensing data. They cover a period between 5.5
and 6.5 years, with on average one measurement every three nights. EROS\,II
observed simultaneously in two non-standard passbands using a dichroic to split
the beam. These passbands are referred to as $V_{\rm E}$ and $R_{\rm E}$, where
the 'E' stands for EROS. These correspond roughly to Johnson-Cousins $V$ and $I$
band (see e.g. de Wit et al. 2003). In their investigation, M02 used the
OGLE\,II data\footnote{Presently only the MACHO and OGLE light curve databases
are publically available through the web.} that cover about 4 years and fully
overlaps in time with the EROS\,II data (see Fig.\,\ref{EOM}). In some cases we
also inspect the publically available MACHO data, whenever the star's light
curve appeared possibly periodic or longer term trends were expected. The MACHO
data set provides the longest time line of the three microlensing projects,
stretching 7 years.  EROS and MACHO data only partially overlap, thus creating
the possibility to construct and investigate light curves stretching nearly 10
years. Illustrating the three baselines, we have plotted in Fig.\,\ref{EOM} all
the available microlensing data for the SMC Be star
\object{OGLE\,004748.79-730519.4}.

\begin{figure}
\includegraphics[height=8cm,width=7cm,angle=90]{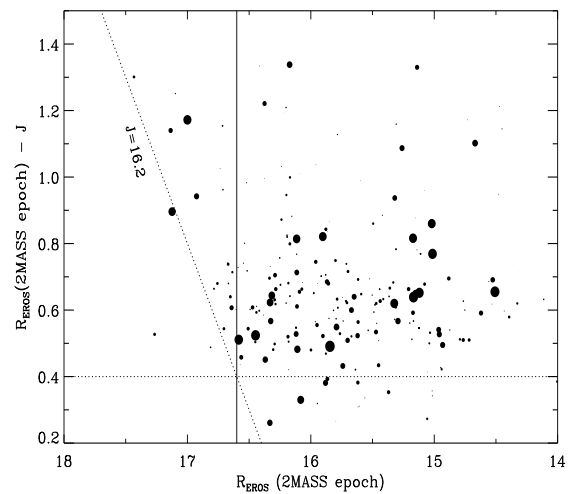}
\caption{Completeness of the 2MASS $J$-band measurements for the selected variable
  stars. Indicated as dotted lines are 2MASS completeness limit and
  approximately the minimum colour excess of the distribution. Stars brighter
  than $R_{\rm E}=16.6^{m}$ do not suffer from 2MASS incompleteness. Sizes of dots is
  proportional to the amplitude of optical variability.}
\label{Complet}
\end{figure}

The 2MASS All Sky data release (Skrutskie et al. 2006\nocite{2006AJ....131.1163S}) was explored to obtain the infrared $JHK$
measurements of the 941 irregularly varying SMC Be stars as reported in M02.  A
cross-correlation between the two datasets resulted in 392 2MASS matches within
a distance of 2 arcsecond, requiring that the 2MASS sources have reliable
photometric measurements in all three near-IR bands. For these objects, 318 EROS
light curve could be extracted.  The Julian Dates of the 2MASS observations are
between 1033.9205 (+2450000) and 1821.7956 (+2450000) and are in principle both
covered by the EROS observations, unfortunately however gaps do exist in the
EROS coverage. The optical magnitude at 2MASS epoch was determined from the red
$R_{\rm E}$ passband (passband 620-920\,nm) observations, allowing a
30-day interval around the 2MASS epoch.  If more than 1 EROS observation is
available in this period, the median of these measurements is taken. The standard
deviation of these measurements is then taken as the uncertainty in the determined
optical magnitude at 2MASS epoch. In this way, for 239 light curves a correspondence
between EROS measurements at the 2MASS epoch is found.

For the large majority of the stars, we are confident that the near-IR 2MASS source
actually corresponds to the variable optical source.  To confirm this, we
searched the optical ($BVI$ bands) OGLE SMC star catalogues (available through
CDS), for objects within 2 arcseconds of the program stars that have $(V-I)$
colours red enough and V-magnitude bright enough to have been picked up by 2MASS
$K$-band observations.  In 11 cases such a bright red star is present in the OGLE
optical SMC star catalogues. We have removed these stars from the sample. With
reference to the variability types of M02, among the selected stars we find 31 of
type 1 (luminosity rise and gradual decline), 33 of type 2 (sudden brightness
jumps in between periods of near constant flux) and 175 of type 4 (light curves
similar to Galactic Be stars) . 

For all of the light curves we determined the minimum and maximum brightness and
the brightness of the object at the 2MASS epoch of observations. The minimum
brightness is assumed to correspond to the photosphere of the Be star. The
photospheres cover a $R_{\rm E}$ magnitude range between $13.9--17.6$. For
the distance of the SMC, these magnitudes would correspond to stars with B
spectral types earlier than B\,5V approximately. The near-IR excess emission flux is
traced by the colour excess defined as $R^{\rm min}_{\rm E}-J$, where the optical
magnitude corresponds to minimum brightness. Given that $R_{\rm E}$ is close
to Johnson $I$-band, we estimate typical photospheric $J$-band magnitudes of the target stars to
be between $14.2-18.1$, as for B-type photospheres one expects $I_{\rm C}-J$ to be
between -0.5 and -0.25. This $J$-band magnitude interval actually brackets the
2MASS $J$-band completeness limit of $\sim 16.2^{m}$. This notwithstanding, the
data is not biased as is shown in Fig.\,\ref{Complet}.  Here the distribution of
near-IR excess is plotted as a function of the EROS magnitude at 2MASS epoch. The
size of each symbol is proportional to the peak-to-peak amplitude of the optical
variability. 

The important thing to notice is that the excess colour is distributed
above a value of 0.4 in $(R_{E}-J)$, independent of optical magnitude. This
indicates that all the stars in our sample have excess colour, which is the main
reason why the sample is not biased. The 2MASS completeness only becomes a
severe effect at optical magnitudes fainter than $\sim16.6^{m}$. In the
following, we thus only consider stars brighter than this limit.

\section{Flux excess in the optical and near-IR}
\label{feon}
\begin{figure}[t]
\includegraphics[height=8cm,width=7cm,angle=90]{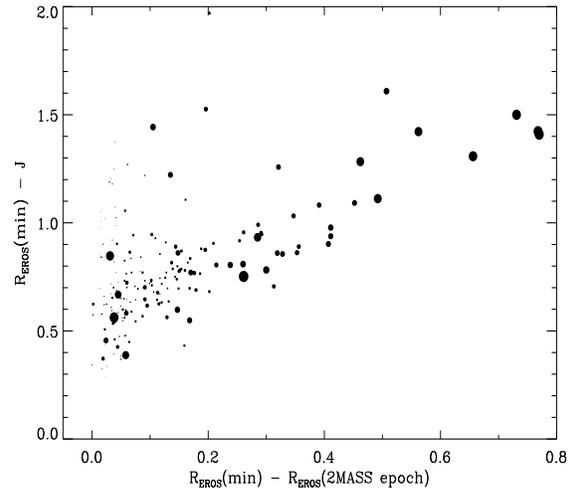}
\caption{The relation between flux excess in the near-IR and the flux
   excess in the optical ($\Delta R$) at 2MASS epoch for all selected 239 SMC Be stars. The size of
   the symbol is proportional to the photometric amplitude of the variability
   measured in the $R_{\rm E}$ passband. The amplitude of variability ranges from 0 to $\sim
   0.9^{m}$.}
\label{ampl}
\end{figure}

The relation  at 2MASS epoch between the infrared excess and the optical flux
excess  for all 239 program stars is plotted in Fig.\,\ref{ampl}.
The infrared excess is evaluated by the $(R_{\rm E}^{\rm min}-J)$ colour and the
optical flux excess by $\Delta R = (R_{\rm E}^{\rm min}-R_{\rm E})$, in which
$R_{\rm E}^{\rm min}$ is the observed minimum brightness of the EROS\,II light curve
over the 5 years of observations. The size of the symbols
corresponds to the $R_{\rm E}$-band amplitude
of variability during the period of EROS\,II observations. The figure shows that
the stars are distributed on an increasingly wider range of $\Delta R$ with increasing near-IR
excess. This suggests that the {\it maximum} amplitude of optical variability a sample
star can display is related to the near-IR excess.

\begin{figure}[t]
\includegraphics[height=8cm,width=7cm,angle=90]{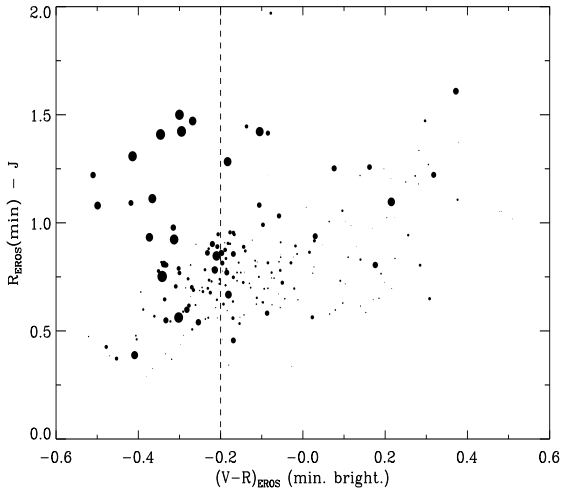}
\caption{The relation between optical colour at minimum brightness of the
  light curve and the near-IR flux excess for all selected SMC Be stars. The size of
   the symbol is proportional to the photometric amplitude of
   variability in the EROS light curve. Generally the small amplitude stars have
   redder optical colours when they have larger near-IR excess. Stars bluer than
   $-0.2$ are represented in Fig.\,\ref{excess}.}
\label{optinear-IR}
\end{figure}

\begin{figure}[t]
\includegraphics[height=8cm,width=7cm,angle=90]{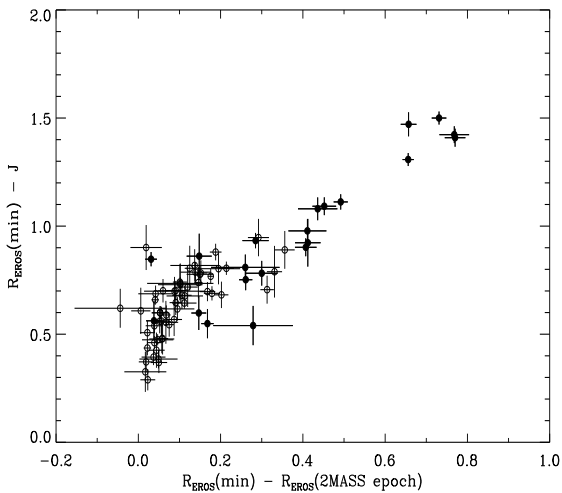}
\caption{The relation between flux excess in the near-IR and the flux
   excess in the optical at the 2MASS epoch of observation. Presented are all 
   stars with an optical colour less than $-0.2$ at brightness minimum. This
   secures that the EROS light curves has covered the epoch when the Be star was
   without CSE. Open symbols have an amplitude of variability less than 0.4.}
\label{excess}
\end{figure}

Fig.\,\ref{ampl} also shows low amplitude stars with large near-IR excess. If indeed these are variable Be stars then
their variability time scale should be longer than the 5 years of continuous
EROS observations. In that case the underlying B-type photosphere has not been
observed and the stars should have optical colours redder than typical B-type
star colours. This is indeed seen in the colour-colour diagram of
Fig.\,\ref{optinear-IR}, relating the near-IR excess with the optical colour at minimum
brightness.  The size of the symbol is again proportional to the amplitude of
variability.  The figure shows that the low amplitude stars (small symbols) are proportionally
redder in the optical $(V-R)_{\rm E}$ at larger near-IR-excess. On the contrary,
the large amplitude stars (large symbols) have generally blue optical colours. It indicates that
for the latter stars the photosphere has been observed and thus that $\Delta R$
in Fig.\,\ref{ampl} is a correct estimate for the optical flux excess at 2MASS epoch.  In
Fig.\,\ref{excess} we again plot therefore the same relation as in Fig.\,\ref{ampl},
but only for those stars with colours $(V-R)_{\rm E} < -0.2$ at brightness
minimum. The relation between the quantities confirms that the optical
irregular variability on observed time scales of 1 to ~1500 days  is correlated with the 
excess flux in the near-IR.

Given the correlations found by K02 between Balmer line emission and variability and
that the variability induces redder colours at brighter phases, our finding here
corroborates the notion that the optical variability on time scales of days to
years is probably best interpreted as variations of the CSE. The observed
maximum amplitude of variability in the optical EROS\,II $R$-band amounts to
$0.9^{m}$, indicating that at peak flux the CSE can contribute about 50\% to the
total optical flux. The distribution of the points in
Figs.\,\ref{ampl} and \ref{optinear-IR} can be attributed to variability time scales
longer than the 5 year period of EROS observations. This long time scale has
inhibited observations of the B-type photosphere of a fraction of the stars, and hence
underestimated the optical flux excess ($\rm \Delta R$) at the 2MASS epoch of observations. Such
a persistent CSE for the low-amplitude stars increases the probability for these
stars to be persistent H$\alpha$ emitters. Cross-correlating all the Be stars in
our sample with the catalogues of SMC H$\alpha$ emitters by Meyssonnier \&
Azzopardi (1993)\nocite{1993A&AS..102..451M} we find a detection rate of 47\%.
The same percentage of H$\alpha$ emitters we obtain for the subsample of
low-amplitude Be stars, not indicating an increased persistence for this
subgroup.

\section{Bi-valued colour-magnitude variation}
\label{colour}

The optical colours of Galactic Be stars are observed to become redder when the
star increases in brightness for the majority of the long-term variable Be stars
(e.g. Moujtahid et al. 1999\nocite{1999A&A...349..151M}; Percy \& Bakos
2001\nocite{2001PASP..113..748P}).  Be star models viewed pole-on by Poeckert
and Marlborough (1978)\nocite{1978ApJS...38..229P} produce this colour-magnitude
relation by scaling the CSE density for a constant outer radius.  The change in
density leads to a change in the emerging bound-free and free-free flux emission
as long as the emission is optically thin. This approach predicts a single
valued relation between colour and magnitude for the B-type star and its varying
CSE.  The opposite trend, i.e. redder optical colour with {\it decreasing}
brightness, is observed but only in a limited number of Galactic Be stars that
are seen under a high inclination (see e.g. Hirata
1982)\nocite{1982IAUS...98...41H}.  The result of the previous section secures
that the optical variability of many of the SMC blue variable stars is due to
ionized material most likely situated in the star's circumstellar
environment. In this section we explore what EROS light curves reveal with
respect to the change in colour as function of the brightness.

\begin{figure*}
\includegraphics[height=17cm,width=7cm,angle=90]{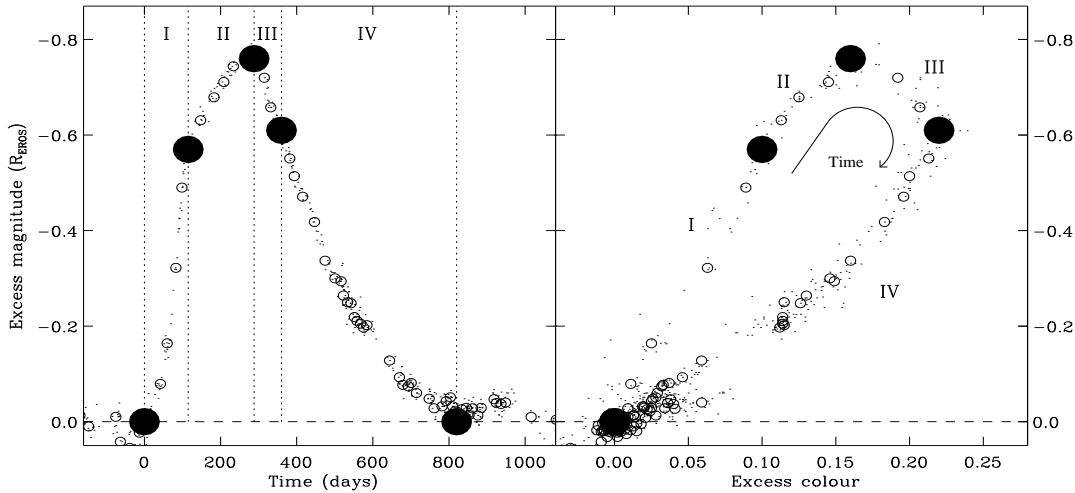}
\caption{Light curve (left panel) and CMD (right panel) of the variation of the star
  OGLE\,005209.92-731820.4. Excess magnitude and excess colour are the 
  observed values minus the magnitude and colour at time = 0 days, i.e. the beginning of
  the flux increase. The small dots represent all measurements, the open
  circles are median values in intervals of ten consecutive measurements.  The
  large black dots mark four specific phases. The excess magnitude temporal
  behaviour produces a loop-like track in the CMD that is traversed in a
  clock-wise sense.
 }
\label{typical}
\end{figure*}

The two panels of Fig.\,\ref{typical} show the light curve and corresponding
colour-magnitude relation of a typical and exemplary variability of one blue
variable SMC star. The colour-magnitude relation is not single valued but double
valued, causing a loop-like relation, and is in contrast with what may be
expected. Zero values in magnitude and colour correspond to the epoch when
minimum flux is observed, presumably the photospheric flux.  The object seems to
have completed a full variability cycle in about 2 years. Fraction wise, some
40\% of the Be stars in our sample that have an amplitude of variability larger
than $0.2^{m}$ in the $R_{\rm E}$-band (101 objects in our sample) trace out a
loop, either during a flux 'burst' or a flux 'dip'. We have chosen this
particular star as an instructive example for the type of colour-magnitude
variability observed among the SMC Be stars.

One can subdivide the observed light and colour curve in four phases (see
Fig.\,\ref{typical}), using the Roman numerals I to IV. As shown by e.g Sabogal et al. (2005)\nocite{2005MNRAS.361.1055S}, 
some 80\% of the Magellanic Cloud blue variables have a shorter flux rising time (phase I and II combined)
than a flux decay time (phase III and IV). 

\begin{itemize}
\item Phase\,I: the initialization phase. It generally corresponds to a relatively
  quick increase in both red and blue flux, causing a redder colour. 
\item Phase\,II: the saturation phase. The flux increase slows down
  significantly. 
\item Phase\,III: the turn-over phase. Both the blue and red flux decline,
  but the blue flux does so faster than the red flux, giving rise to a
  opposite trend between flux and colour, i.e. redder when fainter.
\item Phase IV: the decline phase. The flux decreases and the colour becomes
  bluer. 
\end{itemize}

\begin{figure}
\includegraphics[height=9.5cm,width=8.5cm]{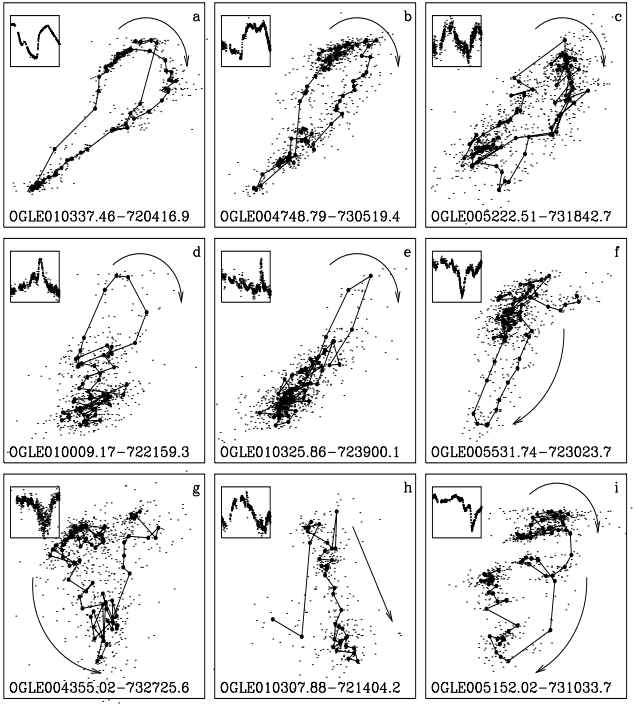}
\caption{Some examples of observed CMDs of SMC Be stars, magnitude on y-axis and colour on
  x-axis. Corresponding ligh curves are in the inset of each panel. The small
  and large dots have the same meaning as in  Fig.\,\ref{typical}. The arrows
  indicate the direction of time. When a loop is observed (101 cases in our
  sample) then in  90\% it is traversed clock-wise.} 
\label{setloop} 
\end{figure}

Interestingly and importantly, these four phases are nearly always observed to
be completed in a clock-wise sense! For 9 instances (10\%) in the subset of the 
sample with an amplitude larger than $0.2^{m}$ we find a loop like feature that
is completed in a counter clock-wise sense, which 
we'll discuss in section Sect.\,\ref{disc}. Examples of this loop
behaviour are illustrated in  
Fig.\,\ref{setloop}. The data in each panel is presented in the same way as in
Fig.\,\ref{typical}, with the exception that the EROS light curve is now shown as
an inset in the upper left corner of each panel. The arrow in each panel
indicates the time consecutive measurements. The first two rows of
Fig.\,\ref{setloop} present SMC Be stars that trace out loop structures in a
clock-wise sence, for various light curve shapes. The first row shows additional examples of
the variability quite similar to the one of Fig.\,\ref{EOM} and
Fig.\,\ref{typical}. These are the so-called bumper stars first reported by the
MACHO collaboration (Cook et al. 1995)\nocite{1995aasp.conf..221C}. On
time scales of years, these Be stars go through a cycle of flux outburst and
subsequent decline back to the presumed photospheric emission. An exception to
the shape of Fig.\,\ref{typical} is presented in panel b where
phases\,I and IV in the CMD partially coincide. The excess flux of Be stars that
vary on somewhat shorter time scale are presented in panel d and e of the
second row. In panel f of Fig.\,\ref{setloop} actual flux decrease and
subsequent recovery to the initial flux level also give rise to two distinct
branches in a CMD. In panel g, an example is shown of loop traversed counter clock-wise.

Finally, other types of colour-magnitude variability observed among the full
sample are illustrated in panels h and i of the third row in Fig.\,\ref{setloop}. 
In these cases there is no indication whatsoever of a loop-like structure, and a
redder when fainter trend is observed (panel h). In the last example we show the
complicated structure of panel i, where for any given colour there can exist as many as three corresponding
magnitudes. For now we refrain from defining various classes or types of
colour-magnitude variability. We conclude with the observation that a bi-valued
(loop-like) colour-magnitude relation seems however a common photometric
phenomenon present among Be stars.

\section{Modeling of the light and colour variations}
\label{model}
The dynamics of Be star disks have been proposed to be viscously dominated,
transporting angular momentum and material outward (Lee et
al. 1991\nocite{1991MNRAS.250..432L}; Porter 1999\nocite{1999A&A...348..512P}).
The kinematics are observed to be consistent with quasi-Keplerian motion (Dachs
et al. 1986\nocite{1986A&AS...63...87D}). Long-term brightness changes can be
understood in terms of an optically thick pseudo photosphere to a more extended
and optically thin envelope (Harmanec 1998\nocite{1998A&A...334..558H}). The
disk loss (transition from Be to B-type star) may occur from the inside out thus
creating a ring e.g. in \object{$o$\,And} (Clark et
al. 2003)\nocite{2003A&A...403..239C}. Ring-like structures are also indicated
by the spectroscopic variability of other Galactic Be stars (Rivinius et
al. 2001\nocite{2001A&A...379..257R}). Below we analyze analytically the effect
on the observed excess emission from emitting ionized material in an outflowing
circumstellar disk. In our simple model the disk is viewed pole-on (in disks 
viewed under an angle, the vertical optical depth will increase by a factor
$({\rm sin}\,i)^{-1}$). Inside-out removal of the disk occurs logically, once the stars
stops losing mass. 

\subsection{An analytical description of the variability}
One can solve analytically for the emerging bf-ff flux emission 
in a time-dependent, outflowing disk (ring) structure by 
applying a geometrically thin disk approximation (i.e. no vertical structure),
which for simplicity is described only by power laws in temperature, velocity, and opening angle
\begin{equation}
  T=T_{0}\,x^{-q},\;\; v=v_{0}\,x^{+p},\;\; \alpha=\alpha_{0}\,x^{+m},
\end{equation}
in which $x=r/R_{*}$.  By applying mass continuity with respect to the mass
loss of the B star, it follows that the density stratification of the disk is then given by
$\rho=\rho_{0}\,x^{n}$, in which the power n equals $-p-m-2$ and $\rho_{0} =
\dot{M}_{*} / 2 \pi R_{*}^2 \alpha_{0} v_{0}$. The absorption coefficients for
free-free and bound-free processes which are proportional to $\lambda^{2}$ and
depend in the same way on the local values of the physical
quantities, and thus a combined expression can be used for their description as given by
Brussaard \& van de Hulst (1962)\nocite{1962RvMP...34..507B}. 
In the Rayleigh-Jeans limit, one can arrive at an analytical expression for the disk
emission by dividing the disk in an optically thick and optically
thin part. For the $\tau>1$ part one finds the expression 
\begin{equation}
\frac{F_{\lambda}}{2 \pi R_{*}^{2}} = \frac{{\rm c_{1}}\,T_{0}\,\lambda^{-4}}{-q+2}\Big\{x_{\tau=1}^{-q+2} -
    x_{\rm in}^{-q+2}\Big\},
\end{equation}
in which ${\rm c_{1}}=2\,k_{b}\,c$. For $\tau<1$ part of the disk
\begin{equation}
\frac{F_{\lambda}}{2 \pi R_{*}^{2}} =
\frac{\epsilon_{0}\,h_{0}\,\lambda^{-2}}{0.5q-2p-m-1}\Big\{ x_{\rm out}^{0.5q-2p-m-1}
- x_{\tau=1}^{0.5q-2p-m-1}\Big\},
\end{equation}
in which $\epsilon_{0}\,h_{0}$ is the product of the emissivity and height of
the disk at the stellar surface. The factor $\epsilon_{0}$ is given by
$c_{ff}\,T^{-0.5}\,\rho_{0}^{2}\,{\rm c_{1}}$, and $h_{0}=R_{*}\,\alpha_{0}$. The
$c_{ff}$ follows from the absorption coefficient for bound-free and free-free processes, see 
e.g. eq.\,1 of Lamers \& Waters (1984). At any wavelength, the inner part
of the disk is more optically thick (in the vertical direction) than the outer
part, because of its $\rho^2$ dependence. Similarly, and at any distance $x$
from the star, the op2tical depth is larger at long wavelength than at short
wavelength, because of its $\lambda^2$ dependence. So for any wavelength, we can
split the disk into an optically thick ring close to the star (if the mass loss
rate is high enough) and an optically thin ring at larger distance.
Having split the emerging flux in two parts, one has to
find the dividing radius for which the vertical optical depth $\tau=1$. This can be done by solving
the appropriate power law expression for the optical depth stratification
\begin{equation}
x_{\tau=1}^\lambda = (\tau_{0}^\lambda\lambda^{2})^{\frac{-1}{1.5q-2p-m-3}},
\end{equation}
The $\tau_{0}$ in this equation is determined from the absorption coefficient
for bf-ff emission, the density, the temperature, and the opening angle, and is
given by $\tau_0^{\lambda}=c_{ff}\,T^{-0.5}\,\rho_{0}^{2}\,h_{0}$. It
represents the vertical optical depth of the disk at the distance $x=1$. Finally the above geometry can be made time dependent by switching on and off
the star's mass loss and solving for the inner and outer rim of the disk given
the power law relation for the outflow velocity (eq.\,1).

The optically thick part (eq.\,2) is governed by q value of the temperature structure 
of the disk as expected. The $\lambda^{-4}$ dependence is a consequence of the
applied Rayleigh-Jeans approximation. The optically thin part is governed by the combined action of the emissivity and
optical depth, determined by the three physical quantities used to approximate
the disk structure, i.e. temperature, radial velocity and opening angle. The
optically thin part has a $\lambda^{-2}$ dependence. The equations describing
optically thin and thick emission thus depend in a different way on the wavelength,
i.e. they have different spectral slopes and thus different colours. The key to the bi-valued
colour-magnitude behaviour is a transition from an optically thick to
an optically thin regime (or vice versa).

\subsection{A simple model for the prediction of variable excess radiation}
\begin{figure}
\includegraphics[height=9.5cm,width=9cm]{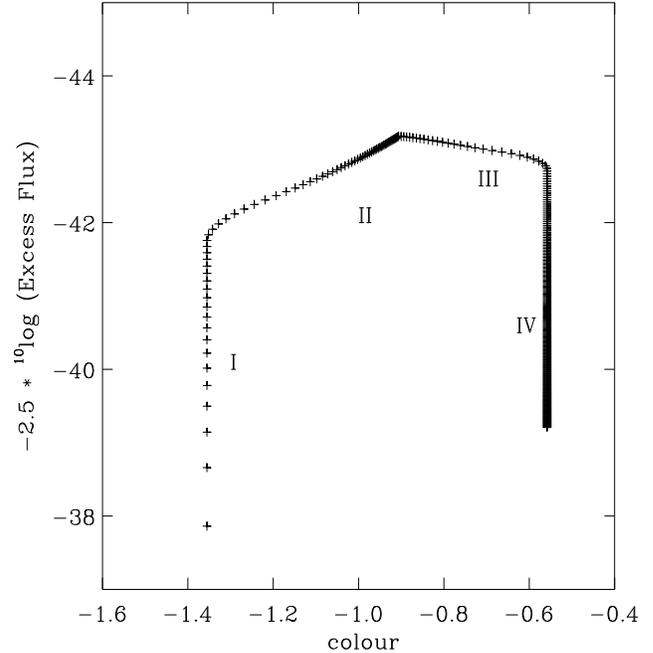}
\caption{Emission from the disk only, as calculated using equations 1--4 for wavelengths $4730\,\AA$
  and $6475\,\AA$. The star loses mass (phase\,I and II) at a rate of
  $10^{-7}\,{\rm M_{\odot}\,yr^{-1}}$ for 1 full year. During phase\,III and IV
  the disk is cleared from the inside out. Crosses mark the evolution every 5
  days. The disk is characterized by $T_{0}=32\,000\,{\rm K}$, $v_{0}=1.0\,{\rm
  km\,s^{-1}}$, and $\alpha_{0}=5^{\circ}$.  Temperature, opening angle and
  radial velocity of the disk are constants during all four phases ($q=p=m=0$).}
\label{pure}
\end{figure}

A calculation using this formalism is presented in Fig.\,\ref{pure}. The figure
shows the logarithm of the emerging disk flux at $\lambda = 6475\,\AA$ against
the colour. The colour is with respect to the emerging disk flux at
$4730\,\AA$. These wavelengths correspond to the effective wavelength of the
EROS dichroics for hot stars. We emphasize that pure disk emission is plotted,
without the stellar radiation. 

The central star has an effective temperature of $40\,000\,{\rm K}$ and a radius
of $R_{*}=7.0\,R_{\odot}$. The temperature of the disk ($T_{0}$ in eq.\,2) is
chosen as 80\% of the stellar effective temperature, i.e. $32\,000\,{\rm
  K}$ and it is chosen to be constant through the disk. In this example, also
the radial outflow velocity of $1.0\,{\rm km\,s^{-1}}$, and the opening angle of
$5^{\circ}$ are constants, and thus we have $q=p=m=0$. The rather high stellar
(and thus disk) temperature is required to comply with the Rayleigh-Jeans
approximation used in the derivation of eqs. 2 and 3.
The disk is produced by a star with a mass-loss rate of $10^{-7}\,{\rm
M_{\odot}\,yr^{-1}}$.  The star continues to lose mass for a period of one year
at this constant rate. This correspond to phases labelled I and II in the CMD of 
Fig.\,\ref{pure}. During this year, the density of the disk at the stellar
surface is always $4.8\, 10^{-10}\,{\rm g\,cm^{-3}}$. Initially (phase\,I) the
disk is optically thick in both the blue and red pass band ($\tau=21$ and
$\tau=44$ respectively). The observed colour of the system is in this case
independent of any disk parameter and only determined by the effective
wavelength of the photometric pass-band, as can be verified from eq.\,2. The
colour of the disk does not change, but the total emerging emission continues to
rise due to the increasing disk radius.

The colour of the disk starts to change in phase\,II, when the disk densities at
the outer disk edge have attained values for which the blue radiation is
optically thin. After sometime also the red radiation becomes optically thin at
the outer disk radius. At the end of phase\,II, the change in colour and
magnitude for each consecutive time step becomes progressively smaller,
asymptotically reaching the colour and emission of an infinitely extended
disk. This corresponds to the peak value of the excess flux. At the beginning of
phase\,III the mass loss is stopped and the disk is cleared from inside out.
Phase\,II and III constitute a transition phase between a fully optically thick
disk and a fully optically thin disk. The disk consists now at both wavelengths
of an optically thick and optically thin part.  In the blue passband the
optically thick disk stretches out to $2.7\,{\rm R_{*}}$ and the red passband
emission is optically thick out to $3.5\,{\rm R_{*}}$.  The disk has a physical
size of $\sim 7.4\,{\rm R_{*}}$. At the outer edge of the disk the optical depths
are $0.05$ and $0.1$ for the blue and red passband respectively. The
termination of the mass loss at phase\,III causes the inner optically thick
parts of the disk to be removed from the inside out, the flux in the red
pass-band to drop, and the overall colour of the disk to approach the colour of
the optically thin disk.  At phase\,IV the optically thick part of the disk has
been completely cleared and the colour corresponds to the ratio of the blue to
red flux in a fully optically thin disk, and determined by eq.\,3. A ring
structure continues to diminish in flux at constant colour, as the density
continues to decrease due to the expansion.

\subsection{An example of modeled variability}
\begin{figure}[t]
\includegraphics[height=8.5cm,width=4cm,angle=90]{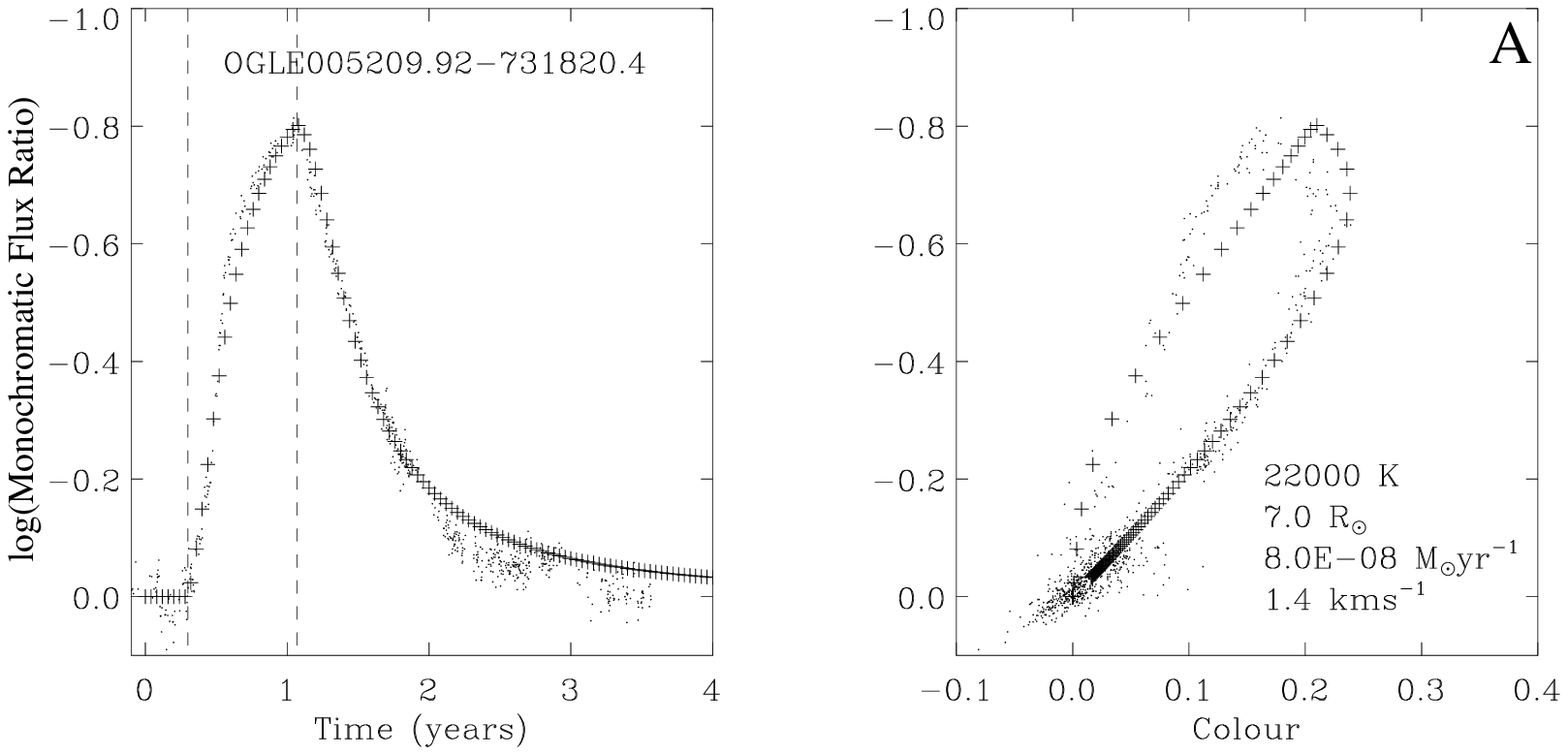}
\includegraphics[height=8.5cm,width=4cm,angle=90]{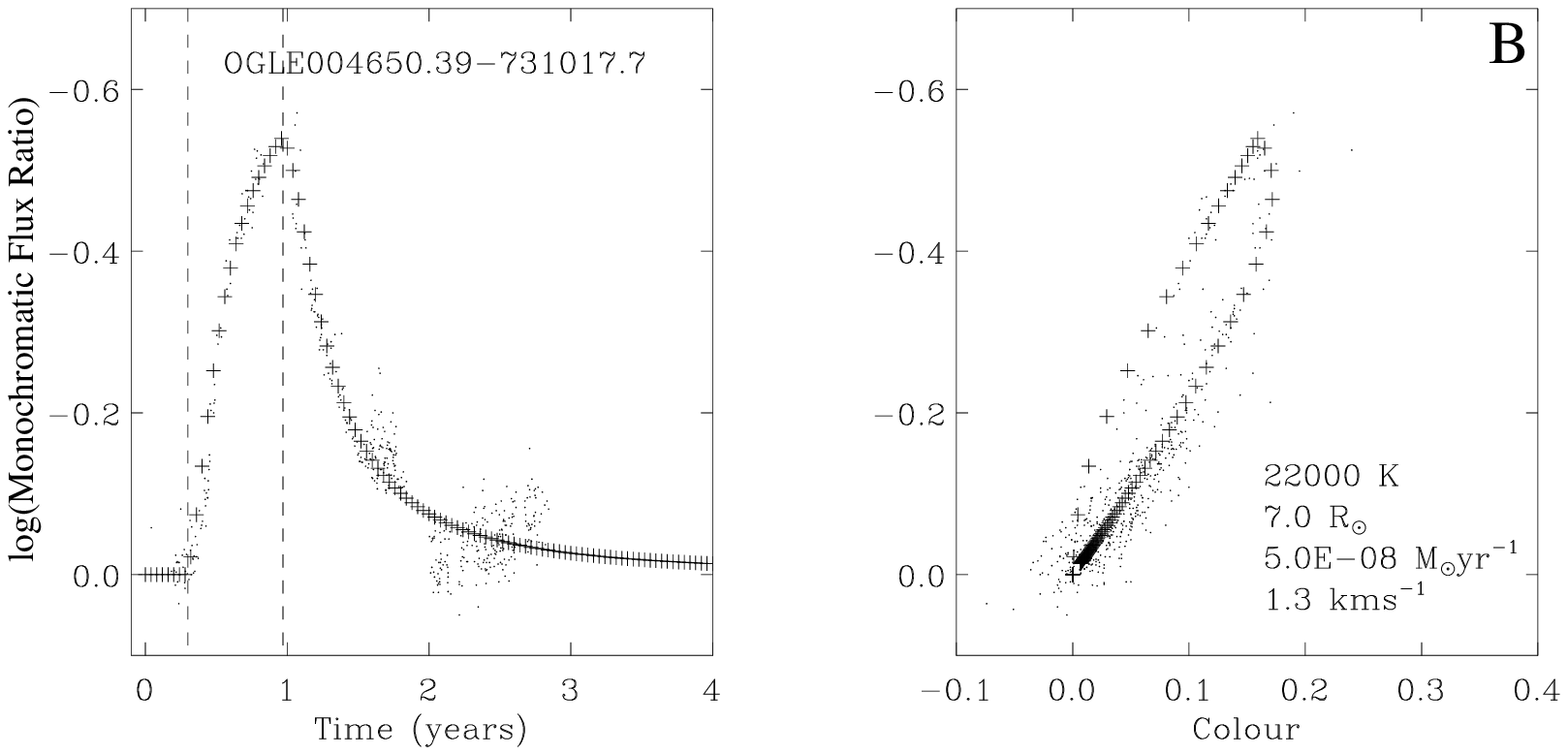}
\includegraphics[height=8.5cm,width=4cm,angle=90]{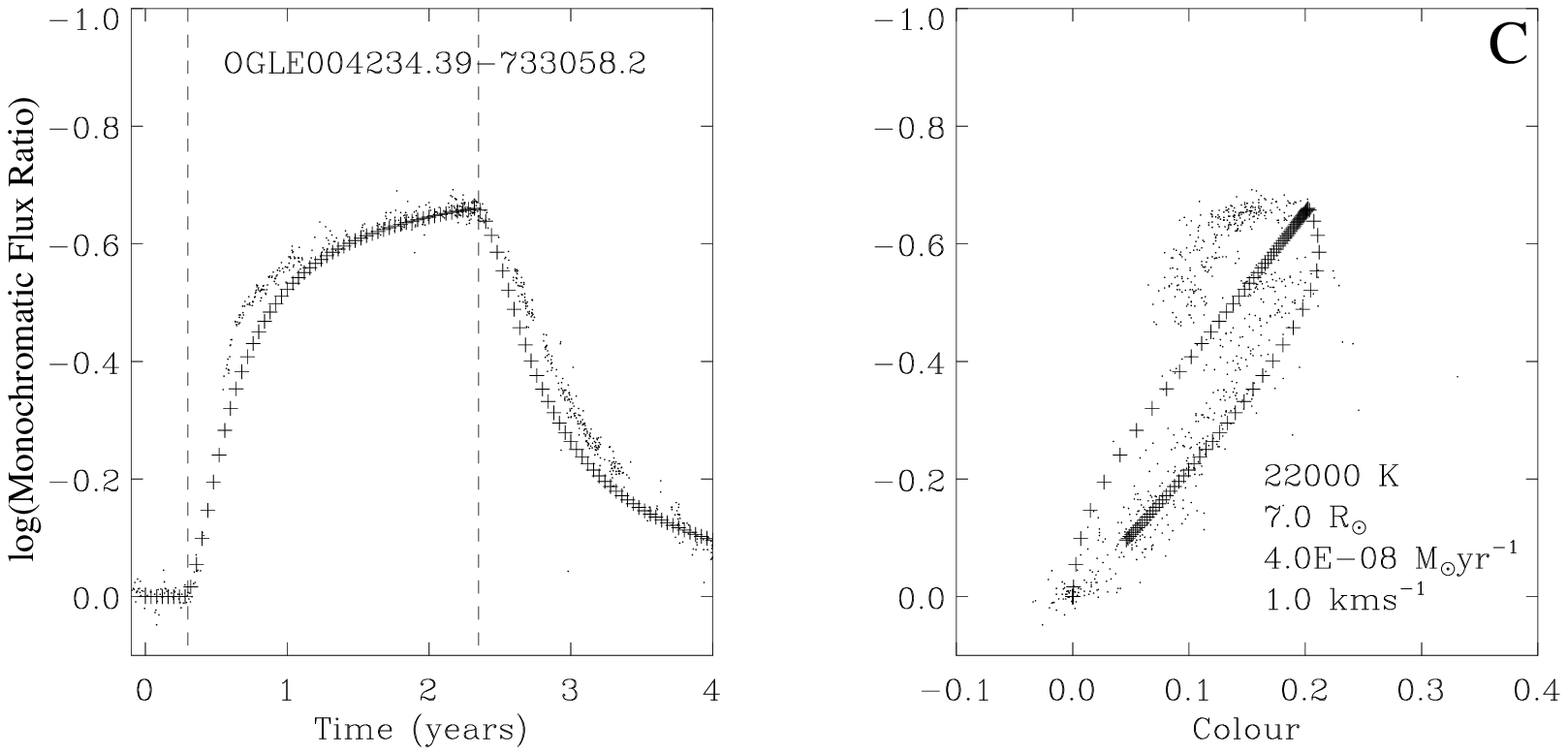}
\caption{Model fits to eruption events of three different SMC Be star (as small dots). 
 The left column are calculated light curves of magnitude against time in
  years. The right column presents the corresponding calculated CMD. The
  parameters of the fit are also given. Temperature, velocity and opening
  angle of the disk are constants. The two dashed lines in the left column
  indicate when the central star was losing mass. See text for further
  explanation of the differences.}
\label{fittry}
\end{figure}

Three illustrations of the model now including a stellar contribution are
presented in the panels of Fig.\,\ref{fittry}.  The calculated fluxes are
compared to observations of three burst events observed in three different Be stars
(small dots). 
The fluxes calculation here is a slight modification to the
analytical model of the previous section. Now the flux is calculated applying
the correct optical depth at each radius, and not the approximation that
the disk consists of a pure optically thick and a pure optically thin part.
Also, now we use the proper expression for the Planck function, and thus we are no
longer restricted to the Rayleigh-Jeans tail of the spectrum. The left column
corresponds to the light curves and the right column to the corresponding
CMDs. In both cases the Y ordinate is chosen to present the $-2.5$ times the
logarithm of the monochromatic flux ratio $Z_{\lambda}=
F_{\lambda}/F^{*}_{\lambda}$ as introduced by Lamers and Waters
(1984)\nocite{1984A&A...136...37L}. The disk temperature at the stellar surface
is chosen to be 0.8 that of the star. For the stars in the sample there is no
spectral information available yet, and we chose therefore a stellar effective
temperature and radius corresponding to the most likely spectral type for the Be
stars in the OGLE sample, i.e. B\,2 (see Mennickent et al. 2002). The CMDs of
the right-hand side panels have as X-ordinate the ratio of the two monochromatic flux ratios
corresponding to the effective wavelengths of the EROS blue and red filters. The
period in which the central star loses mass is marked by the two vertical dashed
lines, and is different for each burst event.

The shape of the light curves are quite similar and we briefly describe this. 
The central B-type star loses mass at a rate of a few times $10^{-8} {\rm
M_{\odot}\,yr^{-1}}$. The material flows out a very modest radial
velocity of about a $1.0\,{\rm km\,s^{-1}}$. During the star's mass losing period, the
excess flux rises steeply as the disk grows from the stellar surface
outward. The colour hardly changes initially, but becomes redder after an
magnitude increase of about $-0.1$. This corresponds to phases\,I and the initial
part of phase\,II in Fig.\,\ref{pure}. The turnover in colour corresponds to
the moment that the blue radiation becomes optically thin at the outer
radius. The flux increase starts to level once the ever increasing outer edge of
the disk has attained densities for which the red radiation also becomes
optically thin. This corresponds to approximately half-way of phase\,II of
Fig.\,\ref{pure}. Once the mass loss is turned off, corresponding to maximum
light, the excess flux drops instantly and quickly, because the optically thick
parts near the stellar surface are the first to disappear; the disk evolves
into a ring. The corresponding colour first becomes redder as {\it relatively}
more optically thick blue flux is removed from the inner edge of the disk
(because the red optically thick part is larger than the blue optically thick
part). Once all optically thick material (blue and red) is removed from the
inner edge of the disk, the colour of the disk follows the track of the
optically thin emission. The exact observed relation between the colour and magnitude
depends not only on the optical thickness of the disk, but also on the stellar
temperature. The reason is that for the optical filters and temperatures of the
star, the Rayleigh-Jeans approximation is no longer valid. Thus by chosing a
disk temperature of 0.8 times the effective temperature of the central star, one
is adding two energy distributions with different spectral slopes. 

All three events were fit by varying the mass-loss rate and the outflow velocity
of the disk material. The outflow velocity, the temperature and the opening
angle of the disk (see eq.\,1) are for each calculation constant. The two panels
of the top row (case A) show the fit to the light curve and the colour-magnitude diagram of
the star already presented in Fig.\,\ref{typical}. In the three cases the
outflow velocity is found to be order $1\,{\rm km\,s^{-1}}$, which is set by
shape of the light curve during the mass losing period. A higher $v_{0}$ has the
effect of (1) making the optically thick part of the disk smaller and (2)
reducing the required time to fill this region with material once the star
starts losing mass. By increasing the mass loss rate, the time scale for filling
in the optically thick part of the disk is the same as for low outflow
velocities (obtaining a similar shape of the flux rise of the light curve) but
for much larger optically thick disk and thus a larger emerging flux.  Thus the
shape of the light curve at the onset of the mass loss of the star can be fit
only with the quoted combination of outflow velocities and mass loss rates. The
disk parameter values quoted in Fig.\,\ref{fittry} depend also on the (unknown)
stellar radius and temperature, and may change by a factor of 2.

The relation between colour and magnitude is not reproduced satisfactorily by
the model. At peak flux the colour is observed to be bluer in case A and C of Fig.\,9.
Note that also in this case the calculation does not reproduce the initial
colour-magnitude behaviour satisfactorily. The two observed branches are much
closer together than the calculated ones. This may indicate the in reality the
mass loss of the star slowly sets in, and does not switch abruptly from zero to
$\rm 10^{-8} M_{\odot}\,yr^{-1}$ as was assumed. Alternatively it may also be
explained if at the initial stages the material has a higher radial velocity or
if the temperature of the disk material is lower. 

\section{Discussion}
\label{disc}
In this paper we have provided evidence that the optical excess emission of an
ensemble of blue variable stars in the Small Magellanic Cloud is correlated with
the near-IR excess emission. The sample contains irregular variable stars, and
no selection was made regarding their type of irregular variability.  It
supports the idea that the blue variable SMC stars, as an ensemble, are stars
showing the Be phenomenon. The maximum optical excess emission found in this
study (nearly 50\% of the continuum) does not seem to be an exceptional feature
for Be stars. Galactic Be type stars have been reported to show similar and even
larger amplitude optical flux out bursts with $\Delta V > 1^{m}$ (e.g. Dachs \&
Wamsteker 1982\nocite{1982A&A...107..240D}; Apparao
1991\nocite{1991ApJ...376..256A}; Mennickent \& Vogt
1991\nocite{1991RMxAA..22..310M}).

The time variable optical excess emission of the SMC blue variable stars show a
bi-valued relation between magnitude and colour. Such a relation between these
two quantities is reported here for the first time. This relation results in a
loop-like track in a CMD. We were able to identify this feature most easily for
stars that have flux outbursts, but they are also observed among stars with flux
decreases. Using an analytic description for the bound-free and free-free
emission of an outflowing disk near a mass-losing B-type star we interpret this
photometric effect in terms of a different time-evolution of the optical depth
for various wavelengths.  The model is phenomenological and is partly based on
recent observations that Be-star disks evolve into ring structures. A
geometrical evolution may occur during the phase of disk dissipation where
material becomes increasingly optically thin from the inside out (e.g. Rivinius
et al. 2001). This may indicate that either material is falling back on the star
or flows out. In the case of the late-type Galactic Be star \object{$o$ And} the
disk seems to be lost from the inside out, but also appears to be rebuilt in a
similar manner, arguing for an outflowing nature of the material (Clark et
al. 2003). In our description the only time-variable parameter is the mass-loss
rate of the central star, that was approximated by only two states (on,
off). Such a mass-loss pattern may be expected if for instance the disk is due
to non-radial pulsations of the central star (e.g. Rivinius et
al. 1998\nocite{1998cvsw.conf..207R}).
Our approach allows us to produce a bi-valued relation between colour and
magnitude, that can fit reasonably well the observed data for objects that show
flux outbursts on time scales between 0.5 and 2 years. In this scenario flux
decreases can be interpreted as a temporarily halt to the mass-loss of the
central star, causing the inner optically thick region to be evacuated.

The colours of Be stars are expected to change continuously, as soon as the star
enters in a mass-losing phase. In only one scenario (within the framework of the
model discussed here) a constant excess colour is observed and that is during
long-lived, constant mass-losing periods in which the circumstellar disk
asymptotically approaches the colour and the excess emission of an infinite
disk. A single relation between colour and magnitude is expected if the disk
material remains optically thin. Multi-valued relation are observed when the
circumstellar environment consists of one or multiple rings. 

\begin{figure}[t]
\includegraphics[height=8cm,width=8cm]{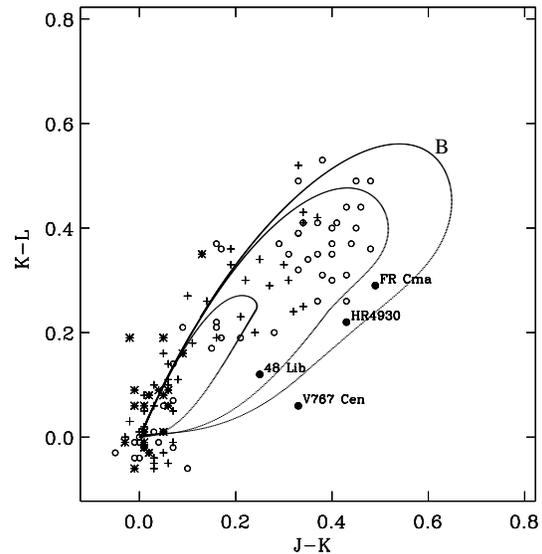}
\caption{The evolution of the near infrared colours for case B of
  Fig.\,\ref{fittry} (outer track), and two cases with a smaller mass loss rate,
  $3\,10^{-8}$ and $\rm 1\,10^{-8}\,M_{\odot}\,yr^{-1}$.  The tracks are traversed
  in a counter clock-wise sense. Symbols represent data taken from Dougherty et
  al. (1991) for early (circles) mid (crosses) and late-type
  (asterisks) Galactic Be stars. During disk creation, one can expect quite blue
  $(K-L)$ colours at given $(J-K)$, as observed for some stars (4 most extreme cases
  are indicated, following Dougherty et al. 1994).}
\label{jkkl}
\end{figure}

The evolution of photometric colours are different from that of static models and may allow to
explain also 'anomalous' near-IR colours as reported in Dougherty et
al. (1994)\nocite{1994A&A...290..609D}, that could not be accounted for using
the static disk model of Waters (1986)\nocite{1986A&A...162..121W}, but can
however be reproduced using an outflowing disk model.  This is shown in
Fig.\,\ref{jkkl}, in which we reproduce Fig.\,8d of Dougherty et al. together
with the evolution of near-IR colours of an outflowing disk/ring structure. As
noted in Dougherty et al.  the four stars indicated by their names have in
particular deviant near-IR colours. These colour are however expected during the
phase of disk build-up. The few stars found in this region is consistent with
the idea that disk creation takes generally much less time than the disk
dispersal.

The bi-valued relation between colour and magnitude results in a loop-like
structure in a CMD. In those cases in the sample for which the amplitude of the
photometric variability is larger than $0.2^{m}$ we could identify a loop in
$40\%$ of the stars. Among these 101 objects the loop is traversed in a
clock-wise sense with time for $90\%$ of the cases. This fact conveys important
information on the physical direction the emitting material is going. Outflowing
material traverses the loop clock-wise, whereas accreting material traverses the
loop counter clock-wise. Simple multi-epoch photometry at two different
passbands can thus establish whether the star is losing or accreting
material. In case of classical Be stars, it can be established whether the
central star reaccretes (part of) its previously expelled material.  If the disk
is (partially) supported by radiation then loss of radiative support will lead
naturally to an accretion phase where the disc falls back on to the star
(e.g. Porter 1999). In our sample of Be stars the re-accretion of matter, using
the loop diagnostic occurs in 10 cases only. This constitutes some 10\% of the
stars showing loops. Thus in general the material in the CS of Be stars (at
least in the SMC) acquires enough angular momentum so as to escape the system
altogether, but the data thus also shows that Be stars in some instances
actually accrete material. Whether these stars are actually classical Be stars
needs to be confirmed. Loop structures in CMDs may thus also be observed in
accreting young stellar objects, in which case the material that is being
accreted changes from optically thin to thick, traversing the loop counter
clock-wise.

\section{Conclusions}
\label{concl}
Irregular optical variability observed in a large ensemble of blue stars in the
Small Magellanic Cloud is due to variations in the amount of Brehmstrahlung,
most likely located in a circumstellar envelope or disk in analogy with
classical Be stars. The CS disk can thus contribute up to 50\% of the total
optical luminosity of the stars, as derived from the amplitude of optical
variability. Using a simple model we have shown that these flux changes can be
interpreted as due to a time variable mass loss of the central star. The disk
material is always outflowing. The outflowing material is observed to produce a
bi-valued colour-magnitude relation (a loop in a colour-magnitude diagram), that
can be ascribed to optical depth effects. The loop is traversed in a clock-wise
sense for outflowing material, but in a anti clock-wise sense for accreted
material. The data is compatible with disks that have a (small) radial outflow
velocity that can mimic a stationary situation. In a small number of instances
the material may actually be falling back on the star.

\begin{acknowledgements}
The authors wish to thank Drs J. Vink and P. Stee for their comments on an
earlier version of the manuscript. This publication makes use of data products
from the Two Micron All Sky Survey, which is a joint project of the University
of Massachusetts and the Infra-red Processing and Analysis Center/California
Institute of Technology, funded by the National Aeronautics and Space
Administration and the National Science Foundation. This paper utilizes public
domain data originally obtained by the MACHO Project, whose work was performed
under the joint auspices of the U.S. Department of Energy, National Nuclear
Security Administration by the University of California, Lawrence Livermore
National Laboratory under contract No. W-7405-Eng-48, the National Science
Foundation through the Center for Particle Astrophysics of the University of
California under cooperative agreement AST-8809616, and the Mount Stromlo and
Siding Spring Observatory, part of the Australian National University.
\end{acknowledgements}


\begin{thebibliography}{30}
\expandafter\ifx\csname natexlab\endcsname\relax\def\natexlab#1{#1}\fi

\bibitem[{{Alcock} {et~al.}(1993){Alcock}, {Akerloff}, {Allsman}, {Axelrod},
  {Bennett}, {Chan}, {Cook}, {Freeman}, {Griest}, {Marshall}, {Park},
  {Perlmutter}, {Peterson}, {Pratt}, {Quinn}, {Rodgers}, {Stubbs}, \&
  {Sutherland}}]{1993Natur.365..621A}
{Alcock}, C., {Akerloff}, C.~W., {Allsman}, R.~A., {et~al.} 1993, \nat, 365,
  621

\bibitem[{{Apparao}(1991)}]{1991ApJ...376..256A}
{Apparao}, K.~M.~V. 1991, \apj, 376, 256

\bibitem[{{Aubourg} {et~al.}(1995){Aubourg}, {Bareyre}, {Brehin}, {Gros}, {de
  Kat}, {Lachieze-Rey}, {Laurent}, {Lesquoy}, {Magneville}, {Milsztajn},
  {Moscoso}, {Queinnec}, {Renault}, {Rich}, {Spiro}, {Vigroux}, {Zylberajch},
  {Ansari}, {Cavalier}, {Moniez}, {Beaulieu}, {Ferlet}, {Grison},
  {Vidal-Madjar}, {Guibert}, {Moreau}, {Tajahmady}, {Maurice}, {Prevot}, \&
  {Gry}}]{1995A&A...301....1A}
{Aubourg}, E., {Bareyre}, P., {Brehin}, S., {et~al.} 1995, \aap, 301, 1+

\bibitem[{{Brussaard} \& {van de Hulst}(1962)}]{1962RvMP...34..507B}
{Brussaard}, P.~J. \& {van de Hulst}, H.~C. 1962, Reviews of Modern Physics,
  34, 507

\bibitem[{{Clark} {et~al.}(2003){Clark}, {Tarasov}, \&
  {Panko}}]{2003A&A...403..239C}
{Clark}, J.~S., {Tarasov}, A.~E., \& {Panko}, E.~A. 2003, \aap, 403, 239

\bibitem[{{Cook} {et~al.}(1995){Cook}, {Alcock}, {Allsman}, {Axelrod},
  {Freeman}, {Peterson}, {Quinn}, {Rodgers}, {Bennett}, {Reimann}, {Griest},
  {Marshall}, {Pratt}, {Stubbs}, {Sutherland}, \&
  {Welch}}]{1995aasp.conf..221C}
{Cook}, K.~H., {Alcock}, C., {Allsman}, H.~A., {et~al.} 1995, in ASP Conf. Ser.
  83: IAU Colloq. 155: Astrophysical Applications of Stellar Pulsation, 221--+

\bibitem[{{Dachs} {et~al.}(1986){Dachs}, {Hanuschik}, {Kaiser}, {Ballereau}, \&
  {Bouchet}}]{1986A&AS...63...87D}
{Dachs}, J., {Hanuschik}, R., {Kaiser}, D., {Ballereau}, D., \& {Bouchet}, P.
  1986, \aaps, 63, 87

\bibitem[{{Dachs} \& {Wamsteker}(1982)}]{1982A&A...107..240D}
{Dachs}, J. \& {Wamsteker}, W. 1982, \aap, 107, 240

\bibitem[{{de Wit} {et~al.}(2003){de Wit}, {Beaulieu}, {Lamers}, {Lesquoy}, \&
  {Marquette}}]{2003A&A...410..199D}
{de Wit}, W.~J., {Beaulieu}, J.-P., {Lamers}, H.~J.~G.~L.~M., {Lesquoy}, E., \&
  {Marquette}, J.-B. 2003, \aap, 410, 199

\bibitem[{{Dougherty} {et~al.}(1994){Dougherty}, {Waters}, {Burki}, {Cote},
  {Cramer}, {van Kerkwijk}, \& {Taylor}}]{1994A&A...290..609D}
{Dougherty}, S.~M., {Waters}, L. B. F.~M., {Burki}, G., {et~al.} 1994, \aap,
  290, 609

\bibitem[{{Harmanec}(1998)}]{1998A&A...334..558H}
{Harmanec}, P. 1998, \aap, 334, 558

\bibitem[{{Hirata}(1982)}]{1982IAUS...98...41H}
{Hirata}, R. 1982, in IAU Symp. 98: Be Stars, 41--43

\bibitem[{{Keller} {et~al.}(2002){Keller}, {Bessell}, {Cook}, {Geha}, \&
  {Syphers}}]{2002AJ....124.2039K}
{Keller}, S.~C., {Bessell}, M.~S., {Cook}, K.~H., {Geha}, M., \& {Syphers}, D.
  2002, \aj, 124, 2039

\bibitem[{{Lamers} {et~al.}(1999){Lamers}, {Beaulieu}, \& {De
  Wit}}]{1999A&A...341..827L}
{Lamers}, H. J. G. L.~M., {Beaulieu}, J.~P., \& {De Wit}, W.~J. 1999, \aap,
  341, 827

\bibitem[{{Lamers} \& {Waters}(1984)}]{1984A&A...136...37L}
{Lamers}, H. J. G. L.~M. \& {Waters}, L. B. F.~M. 1984, \aap, 136, 37

\bibitem[{{Lee} {et~al.}(1991){Lee}, {Osaki}, \& {Saio}}]{1991MNRAS.250..432L}
{Lee}, U., {Osaki}, Y., \& {Saio}, H. 1991, \mnras, 250, 432

\bibitem[{{Mennickent} {et~al.}(2002){Mennickent}, {Pietrzy{\' n}ski},
  {Gieren}, \& {Szewczyk}}]{2002A&A...393..887M}
{Mennickent}, R.~E., {Pietrzy{\' n}ski}, G., {Gieren}, W., \& {Szewczyk}, O.
  2002, \aap, 393, 887

\bibitem[{{Mennickent} {et~al.}(2003){Mennickent}, {Pietrzynski}, \&
  {Gieren}}]{2003ipc..conf...89M}
{Mennickent}, R.~E., {Pietrzynski}, G., \& {Gieren}, W. 2003, in ASP Conf. Ser.
  292: Interplay of Periodic, Cyclic and Stochastic Variability in Selected
  Areas of the H-R Diagram, 89--+

\bibitem[{{Mennickent} \& {Vogt}(1991)}]{1991RMxAA..22..310M}
{Mennickent}, R.~E. \& {Vogt}, N. 1991, Revista Mexicana de Astronomia y
  Astrofisica, 22, 310

\bibitem[{{Meyssonnier} \& {Azzopardi}(1993)}]{1993A&AS..102..451M}
{Meyssonnier}, N. \& {Azzopardi}, M. 1993, \aaps, 102, 451+

\bibitem[{{Moujtahid} {et~al.}(1999){Moujtahid}, {Zorec}, \&
  {Hubert}}]{1999A&A...349..151M}
{Moujtahid}, A., {Zorec}, J., \& {Hubert}, A.~M. 1999, \aap, 349, 151

\bibitem[{{Percy} \& {Bakos}(2001)}]{2001PASP..113..748P}
{Percy}, J.~R. \& {Bakos}, A.~G. 2001, \pasp, 113, 748

\bibitem[{{Poeckert} \& {Marlborough}(1978)}]{1978ApJS...38..229P}
{Poeckert}, R. \& {Marlborough}, J.~M. 1978, \apjs, 38, 229

\bibitem[{{Porter}(1999)}]{1999A&A...348..512P}
{Porter}, J.~M. 1999, \aap, 348, 512

\bibitem[{{Rivinius} {et~al.}(1998){Rivinius}, {Baade}, {Stefl}, \& {et
  al.}}]{1998cvsw.conf..207R}
{Rivinius}, T., {Baade}, D., {Stefl}, S., \& {et al.} 1998, in Cyclical
  Variability in Stellar Winds, ed. L.~{Kaper} \& A.~W. {Fullerton}, 207--+

\bibitem[{{Rivinius} {et~al.}(2001){Rivinius}, {Baade}, {{\v S}tefl}, \&
  {Maintz}}]{2001A&A...379..257R}
{Rivinius}, T., {Baade}, D., {{\v S}tefl}, S., \& {Maintz}, M. 2001, \aap, 379,
  257

\bibitem[{{Sabogal} {et~al.}(2005){Sabogal}, {Mennickent}, {Pietrzy{\'n}ski},
  \& {Gieren}}]{2005MNRAS.361.1055S}
{Sabogal}, B.~E., {Mennickent}, R.~E., {Pietrzy{\'n}ski}, G., \& {Gieren}, W.
  2005, \mnras, 361, 1055

\bibitem[{{Skrutskie} {et~al.}(2006){Skrutskie}, {Cutri}, {Stiening},
  {Weinberg}, {Schneider}, {Carpenter}, {Beichman}, {Capps}, {Chester},
  {Elias}, {Huchra}, {Liebert}, {Lonsdale}, {Monet}, {Price}, {Seitzer},
  {Jarrett}, {Kirkpatrick}, {Gizis}, {Howard}, {Evans}, {Fowler}, {Fullmer},
  {Hurt}, {Light}, {Kopan}, {Marsh}, {McCallon}, {Tam}, {Van Dyk}, \&
  {Wheelock}}]{2006AJ....131.1163S}
{Skrutskie}, M.~F., {Cutri}, R.~M., {Stiening}, R., {et~al.} 2006, \aj, 131,
  1163

\bibitem[{{Udalski} {et~al.}(1997){Udalski}, {Kubiak}, \&
  {Szymanski}}]{1997AcA....47..319U}
{Udalski}, A., {Kubiak}, M., \& {Szymanski}, M. 1997, Acta Astronomica, 47, 319

\bibitem[{{Waters}(1986)}]{1986A&A...162..121W}
{Waters}, L. B. F.~M. 1986, \aap, 162, 121

\end{thebibliography}

\end{document}